\documentclass[showpacs,prd,10pt,twocolumn,superscriptaddress,%
nofootinbib,aps]{revtex4-2}

\usepackage{ae}
\usepackage{bm} 
\usepackage{xcolor}
\usepackage{amsmath}
\usepackage{amssymb}
\usepackage{bm}
\usepackage{amsfonts}
\usepackage{graphicx}
\usepackage{slashed}
\usepackage{wasysym}
\usepackage{units}
\usepackage{hyperref}

\newcommand{\Eqref}[1]{Eq.~\eqref{#1}}

\newcommand{\bk}{\bar{\kappa}}
\newcommand{\bms}{\bar{m}}
\newcommand{\be}{\bar{e}}
\newcommand{\Gprop}{G}
\newcommand{\Nf}{N_{\text{f}}}
\newcommand{\Ni}{N_{\text{i}}}

\usepackage[utf8]{inputenc}

\makeatletter
\def\fps@figure{ht}
\def\fps@table{ht}
\makeatother

\allowdisplaybreaks

\begin{document}

\setlength{\unitlength}{1mm}

\title{Pauli-Term-Induced Fixed Points in $d$-dimensional QED}

\author{Holger Gies}
\affiliation{Theoretisch-Physikalisches Institut, Abbe Center of Photonics, 
Friedrich-Schiller-Universit\"at Jena, Max-Wien-Platz 1, 07743 Jena, Germany}
\affiliation{Helmholtz-Institut Jena, Fr\"obelstieg 3, 07743 Jena, Germany}
\affiliation{GSI Helmholtzzentrum für Schwerionenforschung, Planckstr. 1, 
64291 Darmstadt, Germany} 
\author{Kevin K. K. Tam}
\affiliation{Max Planck School of Photonics, Hans-Kn\"oll-Stra{\ss}e 1, 07745 Jena, Germany}
\affiliation{Theoretisch-Physikalisches Institut, Abbe Center of Photonics, 
Friedrich-Schiller-Universit\"at Jena, Max-Wien-Platz 1, 07743 Jena, Germany}
\author{Jobst Ziebell}
\affiliation{Theoretisch-Physikalisches Institut, Abbe Center of Photonics, 
Friedrich-Schiller-Universit\"at Jena, Max-Wien-Platz 1, 07743 Jena, Germany}

\date{\today}

\begin{abstract}
We explore the fixed-point structure of QED-like theories upon the inclusion 
of a Pauli spin-field coupling. We concentrate on the fate of UV-stable fixed 
points recently discovered in $d=4$ spacetime dimensions upon generalizations 
to lower as well as higher dimensions for an arbitrary number of fermion flavors 
$\Nf$. As an overall trend, we observe that going away from $d=4$ dimensions and 
increasing the flavor number tends to destabilize the non-Gaussian fixed points 
discovered in four spacetime dimensions. A notable exception is a non-Gaussian 
fixed point at finite Pauli spin-field coupling but vanishing gauge coupling, 
which also remains stable down to $d=3$ dimensions and for small flavor 
numbers. This includes also the range of degrees of freedom used in
effective theories of layered condensed-matter systems.  As an application, we 
construct renormalization group trajectories that emanate from the non-Gaussian 
fixed point and approach a long-range regime in the conventional QED${}_3$ 
 universality class that is governed by the 
interacting (quasi) fixed point in the gauge coupling.
\end{abstract}

\maketitle

\section{Introduction}
\label{sec:intro}

The Pauli term, denoting the coupling between the electron spin and the 
electromagnetic field, plays an interesting role in quantum electrodynamics 
(QED): it undergoes finite renormalization \cite{Jackiw:1999qq} while receiving 
contributions from all scales \cite{ZinnJustin:1989mi,Peskin:1995ev}; in the 
effective action, it parameterizes the famous anomalous magnetic moment of the 
electron \cite{Schwinger:1948iu,Schwinger:1951nm} which has been measured and 
computed to an extraordinary precision 
\cite{Laporta:1996mq,Aoyama:2014sxa,Hanneke:2010au}; and from a Wilsonian 
viewpoint, it corresponds to a perturbatively nonrenormalizable dimension-5 
operator and thus has the least possible finite distance to the set of 
renormalizable operators in QED theory space. 

Specifically the last property makes the Pauli term a candidate for a relevant 
interaction in a coupling regime where nonperturbative interactions set in. In 
fact, a recent study \cite{Gies:2020xuh} provides evidence that the observed 
long-range properties of (pure) QED can be extended to high-energy scales along 
renormalization group (RG) trajectories that exhibit a sizable Pauli-term 
contribution. Remarkably, a systematic next-to-leading order
expansion of the effective action features nonperturbative ultraviolet (UV) 
stable fixed points that give rise to a UV-complete version of QED within an 
asymptotic-safety scenario \cite{Weinberg:1976xy,Weinberg:1980gg}. The 
possibility that a sizable Pauli term
could be sufficient for QED to evade the infamous Landau-pole problem 
\cite{Landau:1955} had already been suggested in 
\cite{Djukanovic:2017thn} on the basis of an effective-field-theory study. 

It is important to note that the Pauli-term-induced UV-completion of QED also 
evades the conclusion of QED triviality from previous analyses of QED in the 
strong-coupling regime on the lattice \cite{Gockeler:1997dn} as well as using 
the functional RG \cite{Gies:2004hy}. The reason is that the Pauli term goes 
beyond the chirally invariant subspace of massless QED; moreover, one of the 
non-Gaussian fixed points occurs at vanishing gauge coupling. QED triviality
observed in \cite{Gockeler:1997dn,Gies:2004hy} relies on the observation of 
chiral symmetry breaking induced by a strong gauge coupling which prohibits 
the connection of a strongly coupled high-energy regime to 
the observed phase with a 
light electron. As a caveat, we should mention that the potential of the  
Pauli term to generate a heavy electron mass has not yet been explored. 

The findings of \cite{Gies:2020xuh} serve as a strong inspiration to study the 
Pauli term and the fate of the corresponding nonperturbative fixed points also 
in dimensions larger and smaller than $d=4$. Specifically $d=3$ is a relevant 
case, since QED${}_3$ serves as an effective theory for the 
long-range excitations 
of various layered condensed-matter systems 
\cite{Cortijo:2011aa,Vafek:2013mpa,Franz:2001zz,Franz:2002qy,Herbut:2002wd,
Herbut:2002yq,Tesanovic:2002zz,Mavromatos:2003ss,Herbut:2004ue} including 
graphene and 
cuprate superconductors. In this context, also the dependence of the 
renormalization structure of the theory on the number of fermion flavors $\Nf$ 
is of substantial interest: the question as to whether the long-range 
properties of QED${}_3$ depend on the flavor number and whether quantum phase 
transitions as a function of $\Nf$ exist has a long tradition in the literature 
\cite{Appelquist:1988sr,Nash:1989xx,Pennington:1990bx,Atkinson:1989fp,
Curtis:1992gm,Ebihara:1994wm,Aitchison:1997ua,Appelquist:1999hr,Gusynin:2003ww,
Fischer:2004nq,Kaveh:2004qa,Kubota:2001kk,Hands:2002dv,Hands:2004bh,
Mitra:2006xk,Bashir:2008fk,Bashir:2009fv,Feng:2012we,Grover:2012sp,
Braun:2014wja,Raviv:2014xna,Karthik:2015sza,DiPietro:2015taa,Giombi:2015haa,
Chester:2016wrc,Janssen:2016nrm,Herbut:2016ide,Karthik:2016ppr,Gusynin:2016som,
Gukov:2016tnp,Goswami:2017zts,Janssen:2017eeu,DiPietro:2017kcd,Ihrig:2018ojl,
Benvenuti:2018cwd,Li:2018lyb,Albayrak:2021xtd}.

At the same time, it is interesting to explore the 
features of the system towards higher dimensions: whereas perturbative 
renormalization clearly favors four dimensional spacetime as  
a setting  where interacting field theories of scalar and 
fermionic matter 
with gauge interactions can exist over a wide range of scales, the possibility 
of nonperturbative fixed points appears to loosen this connection between the 
existence of interacting quantum field theories and the observed dimensionality 
of spacetime. Still, various studies find that  evidence for 
nonperturbative UV completions appears to become less robust beyond $d=4$ 
dimensions 
\cite{Gies:2003ic,Morris:2004mg,deForcrand:2010be,Knechtli:2016pph,
Codello:2016muj,Eichhorn:2016hdi,Gracey:2018khg,Fischer:2006fz,Falls:2015qga, 
Gies:2015tca, Eichhorn:2019yzm}.

We start by introducing the model in general dimensions in 
Sect.~\ref{sec:setting}. Section \ref{sec:d4} briefly reviews the results 
derived in \cite{Gies:2020xuh} 
as a reference point for our investigations of lower (Sect.~\ref{sec:d3}) and 
higher (Sect.~\ref{sec:d5}) dimensional spacetimes. Our conclusions are given 
in Sect.~\ref{sec:conc}.

\section{QED${}_d$ with a Pauli term}
\label{sec:setting}

We investigate QED${}_d$, i.e. QED in $d$ spacetime dimensions, with $\Nf$ 
Dirac flavors $\psi^a$ interacting with an electromagnetic field $A_\mu$. In 
addition to the conventional gauge interaction, we include a Pauli spin-field 
coupling already on the level of the bare action. Throughout this work, we use 
Euclidean conventions in which the bare action satisfying Osterwalder-Schrader 
positivity reads 
\begin{equation}
  S \!= \!\int_x \frac{1}{4} F_{\mu\nu} F^{\mu\nu}+ \bar{\psi}^a 
i\slashed{D}[A] \psi^a - i\bms \bar{\psi}^a \psi^a + i \bk
  \bar{\psi}^a \sigma_{\mu\nu} F^{\mu\nu} \psi^a \label{eq:bareEuclS},
\end{equation}
Here the covariant derivative is defined as $D_\mu[A]=\partial_\mu - i\be 
A_\mu$. All mass and coupling parameters are understood to denote bare 
quantities. While the kinetic term is chirally invariant, the mass term $\sim 
\bms$ and the Pauli term $\sim \bk$ break chiral symmetry explicitly. 

In this work, we study the renormalization flow of the couplings and the mass, 
also allowing for wave function renormalizations $Z_{\psi,A}$ that renormalize 
the fields of a Wilsonian-type action of the form of \Eqref{eq:bareEuclS} 
multiplicatively, 
$\psi \to \sqrt{Z_\psi} \psi$, $A \to \sqrt{Z_A} A$. This allows us to define 
the corresponding renormalized couplings. For our present goal of searching for 
fixed points, where the theory becomes (quantum) scale invariant, it is useful 
to introduce dimensionless renormalized quantities:
\begin{equation}
e = \frac{ k^{\frac{d}{2}-2} \be}{Z_\psi\sqrt{Z_A} }, \quad
m = \frac{\bms}{Z_\psi k}, \quad
\kappa =  \frac{k^{\frac{d}{2}-1}\bk}{Z_\psi \sqrt{Z_A}},
\label{eq:dimrencoup}
\end{equation}
where $k$ denotes an RG scale that is used to parameterize RG trajectories in 
the space of couplings. The exponents of $k$ reflect the canonical scaling 
of the corresponding operators. For instance, the gauge  
coupling $e$ is power-counting marginal in $d=4$, relevant in $d<4$ and 
irrelevant in higher 
dimensions. The mass term is a relevant operator in any dimension, whereas
the Pauli term is power-counting irrelevant in all dimensions $d>2$. Note, 
however, that the 
Pauli term, being a dimension-5 operator in $d=4$, has the smallest possible 
distance to marginality in an operator expansion of the action. In addition, it 
is a leading term in a derivative expansion, in which the only other 
dimension-5 term $\sim \bar{\psi} \slashed{D}\slashed{D} \psi$ is 
subleading. 

Even though the canonical scaling fully governs RG (ir-)relevance in the 
perturbative regime, where corrections to scaling can only be logarithmic 
according to Weinberg's theorem \cite{Weinberg:1959nj}, nonperturbative 
phenomena can be characterized by large anomalous dimensions and thus strongly 
affect canonical scaling. In both $d=4$ and $d=3$, the nonperturbative 
phenomenon of chiral symmetry breaking is a prime example for this: at large 
coupling, the anomalous dimensions of the fermionic self-interaction operators 
of the type $\mathcal{O}_{\psi^4} \sim (\bar{\psi} \psi)^2$ can result in 
RG relevance and induce a chiral condensate. This can occur in both $d=4$ 
\cite{Aoki:1996fh,Gies:2005as,Braun:2006jd}, 
where $\mathcal{O}_{\psi^4}$ is a dimension-6 operator, as well as in $d=3$ 
\cite{Kubota:2001kk,Kaveh:2004qa,Herbut:2006cs,Braun:2014wja} where it is a 
dimension-4 operator. These observations put  
an even stronger emphasis on the question of a possible RG relevance of
the Pauli term in the nonperturbative regime, as it is closest to marginality.

We approach the answer to this question by studying the phase diagram of 
QED${}_d$. More specifically, we use the Wetterich equation 
\cite{Wetterich:1992yh,Bonini:1992vh,Ellwanger:1993mw,Morris:1993qb}, 
a functional RG flow equation, and determine the $\beta$ functions of the 
couplings $e$, $m$, and $\kappa$. Using $t=\ln k$ as a flow 
parameter, the $\beta$ functions can be written as
\begin{eqnarray}
\partial_t e &=& \beta_e =  e \left( \frac{d}{2}-2+ \eta_\psi +\frac{\eta_A}{2} 
\right) +\Delta \beta_e, \label{eq:pate}\\
\partial_t m  &=& \beta_m =    -m \left( 1-\eta_\psi\right) + \Delta \beta_m,  
\label{eq:patm}\\
  \partial_t \kappa
  &=& \beta_\kappa = \kappa \left(\frac{d}{2}-1 + \eta_\psi + \frac{\eta_A}{2}\! 
  \right) + \Delta \beta_\kappa,  \label{eq:patkap}
\end{eqnarray}
where $\eta_{\psi,A}$ denote the anomalous dimensions obtained from
\begin{equation}
\eta_\psi=- \partial_t \ln
  Z_\psi , \quad   
\eta_A=- \partial_t \ln Z_A. \label{eq:etas}
\end{equation}
In Eqs.~\eqref{eq:pate}-\eqref{eq:patkap}, we highlighted the dimensional 
scaling terms explicitly, which reflect the canonical scaling exponents already 
displayed in \Eqref{eq:dimrencoup} together with the anomalous dimensions 
$\eta_{\psi,A}$. The last terms $\Delta \beta_{e,m,\kappa}$ abbreviate the 
quantum (loop) contributions. Their explicit forms have been computed in 
\cite{Gies:2020xuh} and are summarized in Appendix \ref{app:loops} to 
next-to-leading order in a systematic expansion scheme of the Wetterich 
equation. Structurally, these fluctuation contributions depend on
\begin{equation}
\Delta\beta=\Delta\beta(e,m,\kappa;\eta_\psi,\eta_A|d,d_\gamma\Nf), 
\label{eq:Delbeta}
\end{equation}
where the anomalous dimensions satisfy algebraic equations also listed in 
App.~\ref{app:loops} and can be expressed as functions of the couplings as 
well. In addition to a parametric dependence on the dimension $d$, the $\beta$ 
functions also depend on the product $d_\gamma \Nf$ counting the number of 
spinor degrees of freedom: in addition to the flavor number $\Nf$, $d_\gamma$ 
denotes the dimensionality of the representation of the Dirac algebra. The 
irreducible representations satisfy $d_\gamma=2^{\lfloor d/2\rfloor}$ which is 
used below unless specified otherwise. 

Rather generally, the $\beta$ functions~\eqref{eq:pate}-\eqref{eq:patkap} are 
not universal, but also depend on the details of the regularization. Even 
perturbatively, only the one- and two-loop coefficient of the marginal coupling 
$e$ in $d=4$ are universal in a mass-independent scheme. Since we include the 
running of the mass and pay attention to threshold effects, we work with a 
standard mass-scale-dependent functional RG scheme. Nevertheless, the existence 
of fixed points of the RG, where all $\beta$ functions vanish, is a universal 
statement. Summarizing the couplings and the $beta$ functions in vector-like 
quantities, $\mathbf{g}=(e,\kappa,m)$, 
$\bm{\beta}=(\beta_e,\beta_\kappa, \beta_m)$, a fixed point $\mathbf{g}^\ast$ 
satisfies $\bm{\beta}(\mathbf{g}=\mathbf{g}^\ast)=0$. In addition, the 
critical exponents $\theta_I$ at a fixed point are also universal. Linearizing the 
$\beta$ functions near a fixed point, the flow is governed by the stability 
matrix $\mathbf{B}$, the eigenvalues of which determine the critical exponents, 
\begin{equation}
\mathbf{B}_{ij} = \frac{\partial \beta_{g_i}}{\partial 
g_j}\Big|_{\mathbf{g}=\mathbf{g}^\ast}, \quad \theta_I= - 
\text{eig}(\mathbf{B}).
\label{eq:stabmat}
\end{equation}
Since we work in an approximation scheme, our results for the universal 
quantities listed in the following also exhibit only approximate universality and 
thus depend on the scheme to some extent. For concrete computations, we use the 
partially linear regularization scheme which is known to be optimized for fast 
convergence to universal results in a class of approximations also used here 
\cite{Litim:2000ci,Litim:2001up}.  

Positive (negative) critical exponents $\theta_i>0$ ($\theta_i<0$) are 
associated with 
relevant (irrelevant) directions. The eigendirections associated with the 
positive exponents span the UV critical surface of trajectories emanating from 
the fixed point. The dimensionality of this surface, and thus the number of 
positive exponents, is equal to the number of physical parameters to be fixed 
in order to render the long-range behavior of the theory fully computable. 
(Eigenvalues $\theta_I=0$ denote marginal directions; here, higher orders 
beyond the linearized flow determine relevance or irrelevance. For 
instance, the QED$_4$ gauge coupling $e$ is marginally irrelevant.)  

In addition to gauge symmetry as a local redundancy, the action has a 
global U($\Nf$) flavor symmetry, whereas an extended 
U$(\Nf)_\text{L}\times$U$(\Nf)_\text{R}$ chiral symmetry is present only in 
the absence of the mass and the Pauli term, $m=0$ and $\kappa=0$. In the general 
case, the action is also invariant under a discrete axial rotation of the 
spinors by an angle of $\pi/2$ and a simultaneous sign flip of $\kappa$ and 
$m$, as well as under charge conjugation and a simultaneous sign flip of $e$ 
and $\kappa$. These latter $\mathbb{Z}_2$ symmetries are also visible on the 
level of the flow equations which remain invariant under $(e,\kappa,m)\to 
(e,-\kappa,-m)$ and $(e,\kappa,m)\to(-e,-\kappa,m)$ as well as combinations 
thereof. Correspondingly, fixed points can exist in these 
sign-flip multiplicities, but, of course, describe one and the same 
universality class. 

In the following, results for fixed-point searches in various dimensions 
and for various fermion degrees of freedom (flavor number, spinor 
representation) are presented. As fixed points at finite couplings are an
inherently nonperturbative phenomenon, it is important to discuss consistency 
criteria in the absence of a generically small control parameter. Our  
systematic approximation scheme represents a combined expansion in operator 
dimension and in derivatives. In this sense, the inclusion of wave function 
renormalizations $Z_{\psi,A}$ already represents a next-to-leading order
contribution. As a quantitative control, we can compare to the leading-order 
result which is obtained by ignoring wave-function-renormalization effects 
in loop terms. In practice, this corresponds to setting $\eta_{\psi,A}=0$ 
inside the loop terms (technically, inside the threshold functions, 
see App.~\ref{app:loops}), but retaining them in the scaling terms displayed 
in Eqs.~\eqref{eq:pate}-\eqref{eq:patkap}. 

We list results only for fixed points where the transition from leading- to 
next-to-leading-order results are quantitatively controlled. This control is 
implemented by demanding that the anomalous dimensions remain sufficiently small, 
$|\eta_{\psi,A}|\lesssim\mathcal{O}(1)$. From a technical viewpoint, the flow 
equations~\eqref{eq:pate}-\eqref{eq:patkap} feature rational functions of 
high order in the couplings on the right-hand sides. Generically, they exhibit 
a large number of fixed-points, most of which do not satisfy our quality 
criteria and are thus considered as artifacts of our approximation. Only a 
small number fulfills the consistency conditions in a remarkably stable manner. 
These are the ones presented in the following sections.

\section{Pauli-term fixed points in $d=4$}
\label{sec:d4}

As a reference point, we start by reviewing the phase diagram of QED$_4$ with a 
Pauli term for $\Nf=1$ as has been found in the literature \cite{Gies:2020xuh}. 
Generalizations to different dimensions and different flavor numbers can 
be well understood by analyzing the similarities and differences to this 
reference case. 

In addition to the Gaussian fixed point $\mathcal{A}$ characterized by 
vanishing couplings, we find two interacting fixed points that satisfy all our 
consistency criteria, cf. Tab.\ref{tbl:fullFixedPoints4}. Both these fixed 
points $\mathcal{B}$ and $\mathcal{C}$ occur at finite Pauli $\kappa$ but 
vanishing gauge coupling. Fixed point $\mathcal{B}$ also occurs at finite mass 
parameter $m$ and has full $\mathbb{Z}_2\times\mathbb{Z}_2$-fold multiplicity, 
whereas for fixed point $\mathcal{C}$ only the charge conjugation multiplicity 
is pertinent.

\begin{table}
	\centering
\begin{tabular}{ccccccccc}
	& $e$ & $\kappa$ & $m$ & multiplicity & $n_{\mathrm{phys}}$ & 
$\theta_{\mathrm{max}}$ & $\eta_\psi$ & $\eta_\mathrm{A}$ \\
	\noalign{\smallskip} \hline \noalign{\smallskip}
	$\mathcal{A:}$ & $0$ & $0$ & $0$ & $-$ & $1$ & $1.00$ & $0.00$ & $0.00$ \\
	$\mathcal{B:}$ & $0$ & $5.09$ & $0.328$ & $\mathbb{Z}_2 \times \mathbb{Z}_2$ & 
$2$ & $3.10$ & $-1.38$ & $0.53$ \\
	$\mathcal{C:}$ & $0$ & $3.82$ & $0$ & $\mathbb{Z}_2$ & $3$ & $2.25$ & $-1.00$ 
& $0.00$ \\
\end{tabular}
\caption{Fixed points of $d=4$ dimensional QED.}
\label{tbl:fullFixedPoints4}
\end{table}

\begin{figure}[t]
	\includegraphics[width=0.45\textwidth]{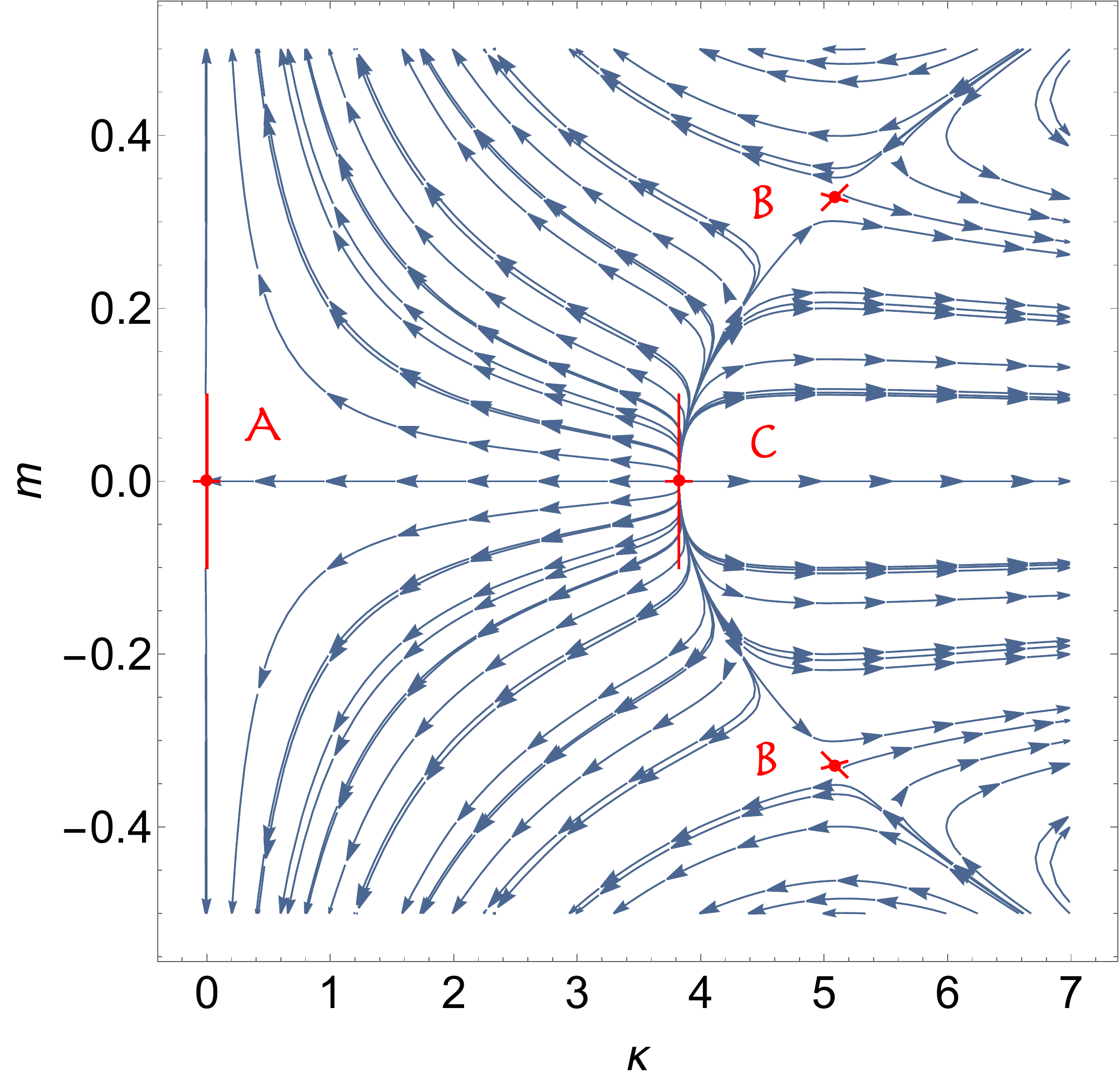}
	\caption{Phase diagram for $d=4$, $\Nf=1$ in the $(\kappa,m)$ plane at 
vanishing $e=0$ exhibiting the Gaussian fixed point $\mathcal{A}$, and the 
non-Gaussian fixed points $\mathcal{B}$ and $\mathcal{C}$ (a $\mathbb{Z}_2$ 
reflection of $\mathcal{B}$ is also visible). Arrows denote the renormalization group 
flow towards the IR. The massive phase where the dimensionless mass $m$ scales as 
$m\sim 1/k$ to large values is accessible from all fixed points.}
	\label{fig:StreamPlotAtEEqualTo0d4}
\end{figure}

At the Gaussian fixed point $\mathcal{A}$, only the mass is a relevant 
direction. At fixed point $\mathcal{B}$, the direction towards finite 
gauge coupling also represents a relevant direction. Fixed point $\mathcal{C}$ 
features 3 relevant directions, implying that all couplings in the action 
correspond to physical parameters that define the long-range behavior of the 
system. The flow towards the long-range IR is visualized in the phase diagram 
of the $(\kappa,m)$ plane at $e=0$ in Fig.~\ref{fig:StreamPlotAtEEqualTo0d4}. 
Note that the rapid flow near the $m$ axis towards large values of $m$ reflects 
the fact that $m$ denotes a dimensionless mass parameter increasing as 
$m\sim1/k$ for $k\to0$ if the physical mass approaches a finite value.

As discussed in \cite{Gies:2020xuh}, UV-complete trajectories that agree with 
the observed long-range behavior of pure QED$_4$ can be constructed with fixed 
point $\mathcal{C}$ as a UV fixed point. Even though UV-complete trajectories 
emanating from $\mathcal{B}$, of course, also exist, their long-range behavior 
is characterized by very large values of the 
anomalous magnetic 
moment of the electron, incompatible with observations.  

Beyond arguments of physical compatibility (ultimately, pure QED is only a part 
of the electroweak sector), a second glance at the non-Gaussian 
fixed points further reveals more subtle differences: at fixed point 
$\mathcal{C}$, the fermion anomalous dimension is exactly $\eta_\psi=-1$ which 
is the value needed to render the Pauli coupling power-counting 
marginal. The fermionic scaling dimension at the fixed point becomes equivalent 
to that of a scalar field for this value. In order to quantify how 
nonperturbative the system is at the fixed points, let us introduce the quantity
\begin{equation}
 \alpha_\kappa= \frac{\kappa^2}{4\pi}, 
  \label{eq:alphakap}
\end{equation}
in analogy to the fine-structure constant $\alpha=\frac{e^2}{4\pi}$. At the 
fixed point $\mathcal{C}$, we observe that $\alpha_\kappa^\ast\simeq 1.16$, 
whereas $\alpha_\kappa^\ast\simeq 2.06$ at fixed point $\mathcal{B}$. This, 
together with the fact that the anomalous dimensions are larger, 
indicates that fixed point $\mathcal{B}$ is in a significantly deeper 
nonperturbative region. 

Let us now go beyond the literature results of \cite{Gies:2020xuh} and study 
the flavor number $\Nf$ dependence of the phase diagram for the irreducible 
representation $d_\gamma=4$. Since we have $e=0=m$ at the fixed point 
$\mathcal{C}$, the flow equation for the remaining coupling $\kappa$ simplifies 
considerably, yielding
\begin{equation}
 \partial_t \kappa|_{e,m=0}= (1+\eta_\psi) \kappa, \quad \eta_\psi=- 
\frac{3}{5\pi^2} \frac{\kappa^2}{1- \frac{3}{40\pi^2}\kappa^2}.
\label{eq:patkaponly}
\end{equation}
We observe that $\eta_\psi=-1$ is a requirement for the existence of a 
fixed-point at the present level of approximation which is indeed satisfied for 
$\mathcal{C}$. Furthermore, the flow and thus also its fixed points are 
completely independent of $\Nf$. The same statement also holds for the 
stability matrix $\mathbf{B}$ at the fixed point and thus also for the critical 
exponents. We conclude that the results for fixed point $\mathcal{C}$ in 
Tab.~\ref{tbl:fullFixedPoints4} persist for any value of $\Nf$. 

\begin{figure}[t]
\includegraphics[width=0.45\textwidth]{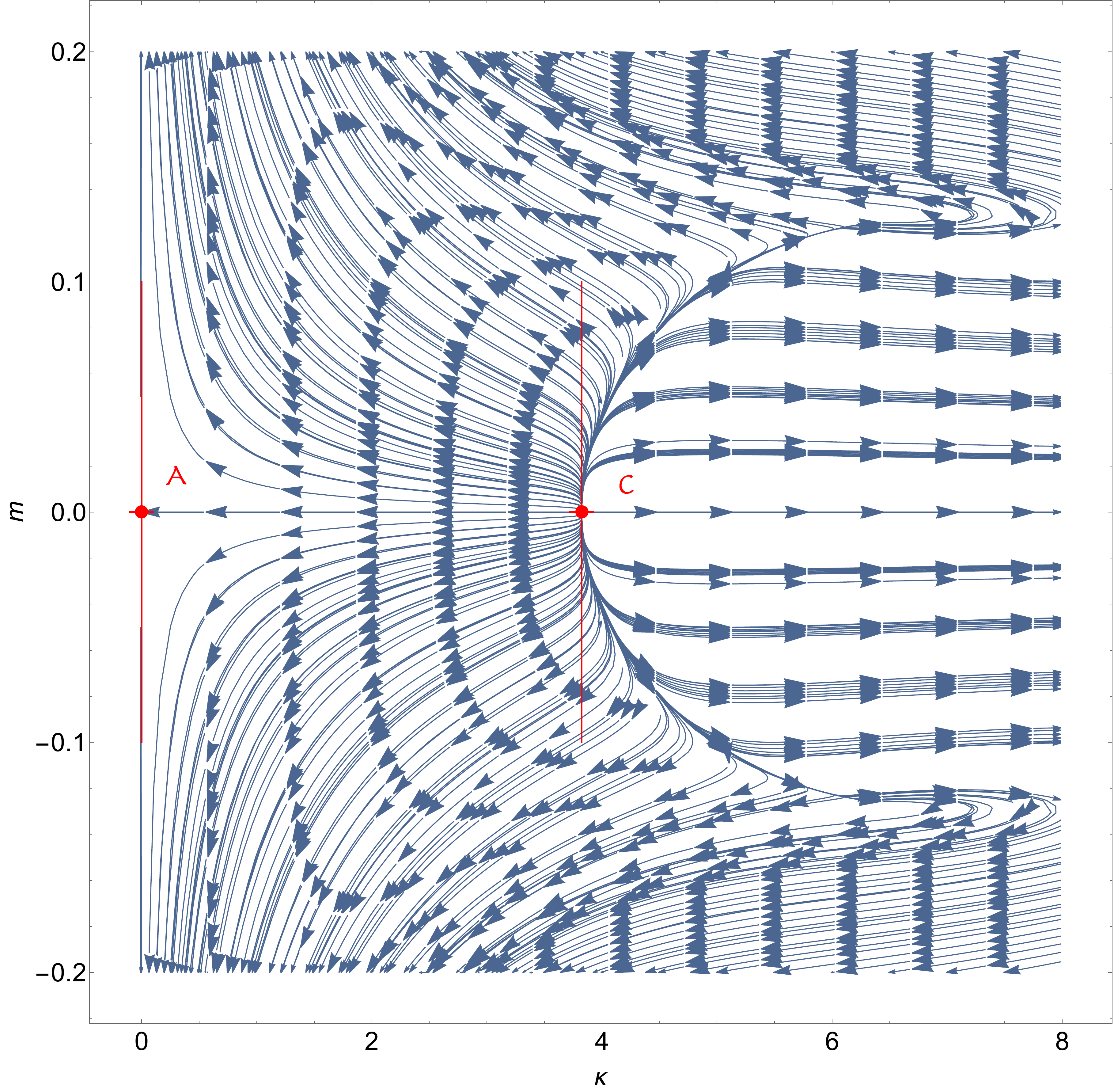}
	\caption{Phase diagram  for $d=4$, $\Nf=19$ in the $(\kappa,m)$ plane at 
vanishing $e=0$. In comparison to $\Nf=1$, the fixed point $\mathcal{B}$ has 
disappeared as a consequence of a fixed point collision, whereas the
Gaussian fixed point $\mathcal{A}$ and the non-Gaussian fixed 
point $\mathcal{C}$ persist for all values of $\Nf$.}
	\label{fig:StreamPlotAtEEqualTo0d4Nf19}
\end{figure}

This is different for fixed point $\mathcal{B}$ where both fixed point position 
and critical exponents do depend on $\Nf$. We observe that $\kappa^\ast$ at the 
fixed point increases with $\Nf$ whereas $m^\ast$ decreases. At a critical 
value of the flavor number $\Nf\simeq 18.50$, fixed point $\mathcal{B}$ collides with another fixed point and 
then disappears into the complex plane. 
This other fixed point comes from even larger values of the coupling and 
belongs to those that do not satisfy our consistency conditions at $\Nf=1$. In 
fact, near the collision, fixed point $\mathcal{B}$ no longer satisfies our 
consistency conditions. For instance, at $\Nf=18$, we have $\eta_\psi\simeq 
-2.65$, $\eta_A=3.24$, and $\kappa^\ast \simeq6.13$, implying $\alpha_\kappa 
\simeq 3$. A phase diagram for $\Nf=19$ where $\mathcal{B}$ has disappeared is 
shown in Fig.~\ref{fig:StreamPlotAtEEqualTo0d4Nf19}.

As a summary of the $d=4$ dimensional theory, we find that fixed point 
$\mathcal{C}$ exists for all values of $\Nf$, remains quantitatively stable and 
supports UV-complete trajectories that can be matched to the physical 
long-range observables. By contrast, fixed point $\mathcal{B}$ becomes less 
consistent with increasing $\Nf$ and disappears for large $\Nf$ in a 
fixed-point collision. Since this collision occurs in a regime which is not 
well controlled in our approximation, we anticipate that the critical flavor 
number estimated here as $\Nf\simeq 18.50$ can undergo large 
corrections upon improvements of the approximation.

\section{Pauli-term fixed points in lower dimensions}
\label{sec:d3}

In spacetime dimensions lower than four, $d<4$, the RG flow of the couplings 
exhibits both qualitative as well as quantitative changes. Quantitatively, we 
observe in Eqs.~\eqref{eq:DeltaBetaE}-\eqref{eq:etaA} that a sizable number of terms have a prefactor 
$(d-4)$ and thus contribute only away from $d=4$. 

Qualitatively, a major change occurs, since the scaling term of the gauge 
coupling flow \Eqref{eq:pate} now contributes to linear order in $e$, i.e., 
$\partial_t e = -\frac{1}{2} e + \mathcal{O}(e^3,\dots)$. Provided that the loop 
terms of higher order are sufficiently positive, a new fixed point 
$\mathcal{D}$ emerges on the positive $e$ axis (with a corresponding 
$\mathbb{Z}_2$ reflection at negative values of $e$). The occurrence of this 
fixed point $\mathcal{D}$ is well known in the literature 
\cite{Pisarski:1984dj,Stam:1985tb,Appelquist:1988sr}. It is infrared (IR) 
attractive in the coupling 
flow in contrast to the Gaussian fixed point $\mathcal{A}$ at which the 
gauge coupling $e$ now parameterizes a relevant IR-repulsive interaction. 

The fixed-point value $e^\ast$ at $\mathcal{D}$ is decisive for the long-range 
behavior of the lower-dimensional QED. If $e^\ast$ is sufficiently weak, chiral 
symmetry can persist: along a trajectory emanating from the Gaussian fixed 
point $\mathcal{A}$, the system remains massless and the theory can be IR 
conformal. For such trajectories, $e^\ast$ at $\mathcal{D}$ represents the 
maximum long-range coupling strength of the theory. 

If this maximum coupling is sufficiently large, it has the potential to trigger 
chiral symmetry breaking and the formation of chiral condensates. From an RG 
picture, symmetry breaking can proceed via induced fermionic self-interactions 
becomes relevant \cite{Kubota:2001kk,Kaveh:2004qa,
Braun:2014wja}, rendering fixed point $\mathcal{D}$ unstable through a 
fixed-point collision; in this case, $\mathcal{D}$ represents a quasi fixed 
point that exists only for a finite range of the RG flow and disappears in 
the deep IR. Whether this occurs or not in 
the physically relevant case $d=3$, and for which range of $\Nf$, has been a 
major research thread in the past decades 
\cite{Appelquist:1988sr,Nash:1989xx,Pennington:1990bx,Atkinson:1989fp,
Curtis:1992gm,Ebihara:1994wm,Aitchison:1997ua,Appelquist:1999hr,Gusynin:2003ww,
Fischer:2004nq,Kaveh:2004qa,Kubota:2001kk,Hands:2002dv,Hands:2004bh,
Mitra:2006xk,Bashir:2008fk,Bashir:2009fv,Feng:2012we,Grover:2012sp,
Braun:2014wja,Raviv:2014xna,Karthik:2015sza,DiPietro:2015taa,Giombi:2015haa,
Chester:2016wrc,Janssen:2016nrm,Herbut:2016ide,Karthik:2016ppr,Gusynin:2016som,
Gukov:2016tnp,Goswami:2017zts,Janssen:2017eeu,DiPietro:2017kcd,Ihrig:2018ojl,
Benvenuti:2018cwd,Li:2018lyb,Albayrak:2021xtd} with different 
methods yielding different answers.

\begin{figure}[t]
\includegraphics[width=0.45\textwidth]{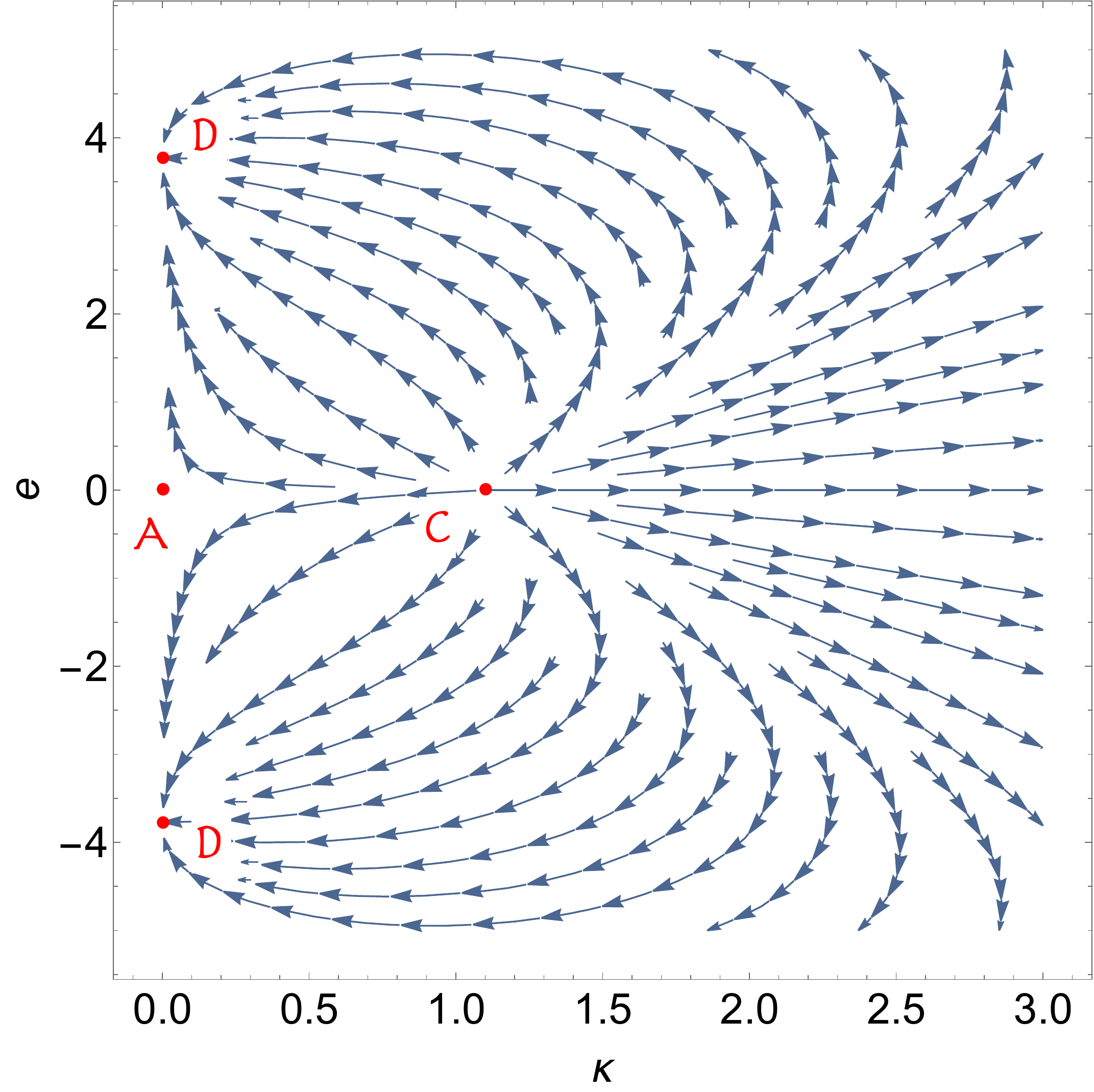}
	\caption{Phase diagram for $d=3$, $\Nf=1$ in the $(e,\kappa)$ plane at 
vanishing $m=0$ for the irreducible representation 
$d_\gamma=2$. The IR-attractive fixed point $\mathcal{D}$ is 
characteristic for lower-dimensional QED. It can characterize a 
chirally invariant long-range conformal phase or--if sufficiently strongly 
coupled--trigger dynamical chiral symmetry breaking. This fixed point also 
remains IR attractive in the direction of the Pauli coupling.}
	\label{fig:StreamPlotAtMEqualTo0d3Nf1}
\end{figure}

While the present work has nothing to add to the question of chiral symmetry 
breaking of the system at or near fixed point $\mathcal{D}$, let us now study 
how the Pauli coupling complements the phase diagram away from the chirally 
symmetric subspace. In fact, we observe the existence of the Pauli-coupling 
fixed point $\mathcal{C}$ for all lower dimensions in between $2<d<4$ with 
qualitative properties similar to $d=4$ for sufficiently small flavor numbers. 
For the relevant case 
of $d=3$ and $\Nf=1$ irreducible Dirac flavors with 
$d_\gamma=2$, both non-Gaussian fixed points can be seen in the 
$(e,\kappa)$ plane at $m=0$ in Fig.~\ref{fig:StreamPlotAtMEqualTo0d3Nf1}. This 
phase diagram illustrates that fixed point $\mathcal{D}$ remains IR attractive 
also in the direction of the chiral symmetry breaking Pauli 
coupling; by contrast, a mass-type perturbation remains relevant. We also 
observe that fixed point $\mathcal{C}$ is fully repulsive. Quantitatively, it 
is interesting to see that the fixed-point value of the Pauli coupling at 
$\mathcal{C}$ decreases with decreasing dimension; e.g., we have 
$\alpha_\kappa\simeq 0.1$ at $d=3$.

\begin{figure}[t]
\includegraphics[width=0.45\textwidth]{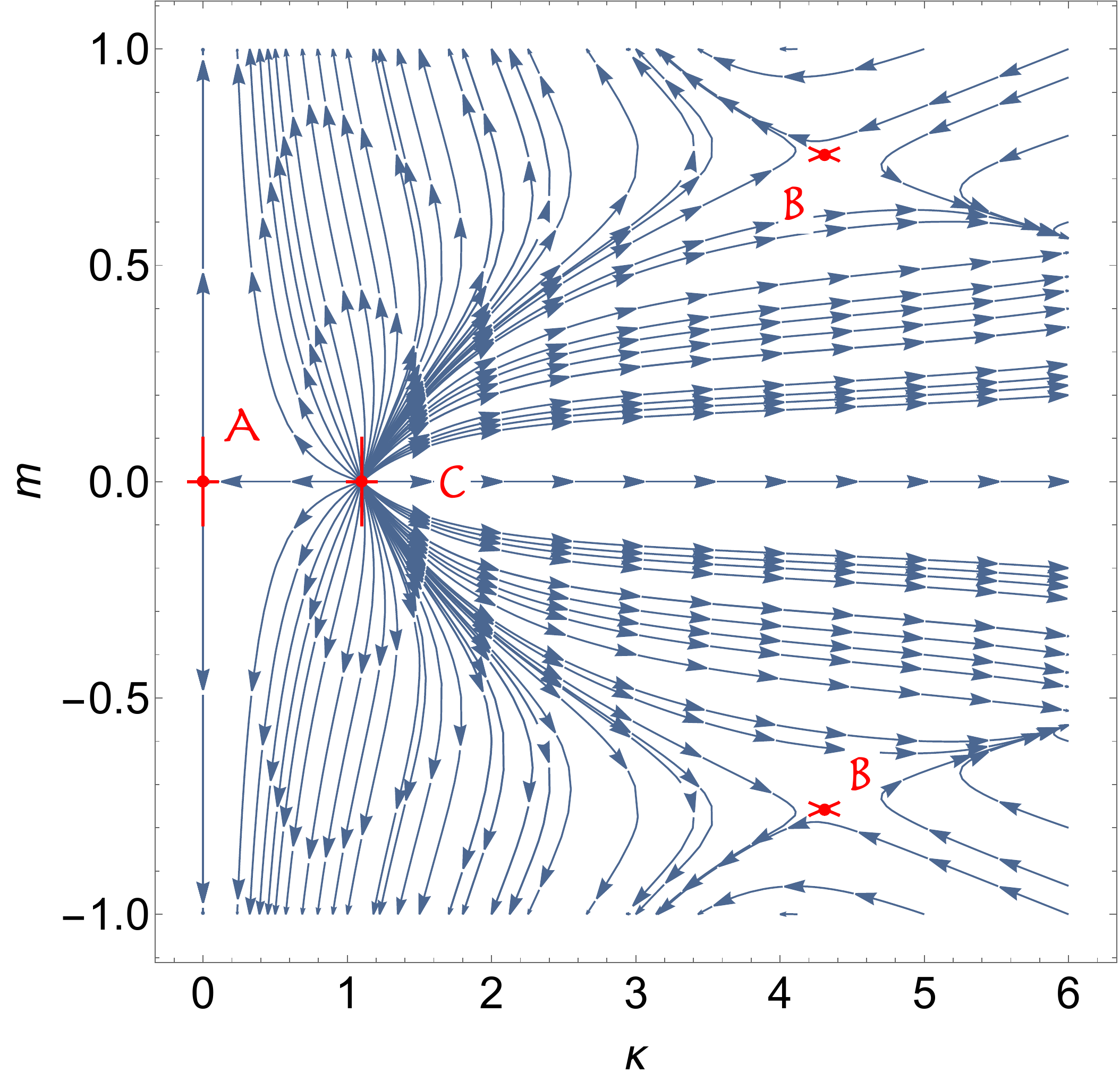}
	\caption{Phase diagram  for $d=3$, $\Nf=1$ in the $(\kappa,m)$ plane at 
vanishing $e=0$ for the irreducible representation 
$d_\gamma=2$. In comparison to 
Fig.~\ref{fig:StreamPlotAtEEqualTo0d4} in $d=4$, the fixed point $\mathcal{C}$ 
is more weakly coupled.}
	\label{fig:StreamPlotAtEEqualTo0d3Nf1}
\end{figure}

Quantitative results for the fixed points for $d=3$ and $\Nf=1$ 
irreducible flavors are listed in 
Tab.~\ref{tbl:fullFixedPointsd3Nf1dg2}. In addition, the non-Gaussian fixed point 
$\mathcal{B}$ is present at smaller values of $\kappa^\ast$ but larger mass 
parameter values $m^\ast$ in comparison to $d=4$, cf. 
Fig.~~\ref{fig:StreamPlotAtEEqualTo0d3Nf1}. More critically, the fixed point 
exhibits rather large anomalous dimensions and thus no longer fully meets 
our consistency criteria in contrast to fixed point $\mathcal{C}$. In fact, 
fixed point $\mathcal{B}$ undergoes a fixed-point collision with an inconsistent 
fixed point somewhat below $d\simeq 2.9$ and thus disappears from the spectrum. 
We take this as an indication that fixed point $\mathcal{B}$ is likely to also 
be an artifact in $d=3$.

\begin{table}[t]
	\centering
\begin{tabular}{ccccccccc}
	& $e$ & $\kappa$ & $m$ & multiplicity & $n_{\mathrm{phys}}$ & 
$\theta_{\mathrm{max}}$ & $\eta_\psi$ & $\eta_\mathrm{A}$ \\
	\noalign{\smallskip} \hline \noalign{\smallskip}
	$\mathcal{A:}$ & $0$ & $0$ & $0$ & $-$ & $2$ & $1.00$ & $0.00$ & $0.00$ \\
	$\mathcal{B:}$ & $0$ & $4.31$ & $0.757$ & $\mathbb{Z}_2 \times \mathbb{Z}_2$ 
& 
$2$ & $1.69$ & $-1.84$ & $2.42$ \\
	$\mathcal{C:}$ & $0$ & $1.10$ & $0$ & $\mathbb{Z}_2$ & $3$ & $1.03$ & $-0.426$ 
& $0.121$ \\
	$\mathcal{D:}$ & $3.77$ & $0$ & $0$ & $\mathbb{Z}_2$ & $1$ & $2.98$ & $-0.571$ 
& $0.885$ \\
\end{tabular}
\caption{Fixed points of $d=3$ dimensional QED for $\Nf=1$ 
irreducible flavors ($d_\gamma=2$). Potential dynamical 
chiral symmetry breaking triggered near the strong-coupling IR fixed point 
$\mathcal{D}$ is not accounted for.}
\label{tbl:fullFixedPointsd3Nf1dg2}
\end{table}

For completeness, we add that towards even lower dimensions, e.g., at $d\lesssim 
2.28$, further $\mathcal{B}$-type fixed points reappear again 
satisfying our consistency criteria with $\eta_{\psi,A}\lesssim 
\mathcal{O}(1)$. This is reminiscent of the occurrence of multicritical models 
for $d<3$ with an increasing number of models towards $d\to2$ in scalar 
O($N$) theories 
\cite{ODwyer:2007brp,Codello:2017hhh,Codello:2017qek,Martini:2018ska}. A proper 
resolution of such 
fixed points, however, requires a larger set of operators in the truncated 
theory space.

\begin{figure}[t]
\includegraphics[width=0.45\textwidth]{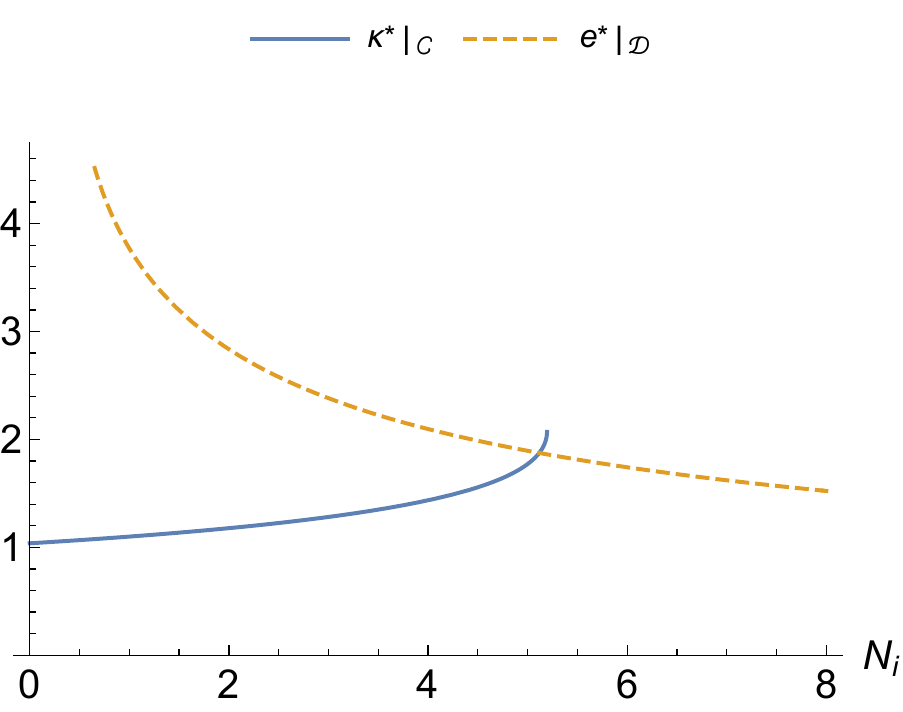}
	\caption{Fixed-point values $\kappa^\ast$ at fixed point $\mathcal{C}$ 
and $e^\ast$ at fixed point $\mathcal{D}$ as a function of the irreducible 
flavor number $\Ni$ in $d=3$ spacetime dimensions. Fixed point $\mathcal{C}$ 
disappears in a fixed-point collision slightly above $\Ni=5.1$.}
	\label{fig:EAtDandKappaAtCd3}
\end{figure}

Let us now study the theory for larger flavor numbers. For concreteness, we 
restrict ourselves to the relevant case of $d=3$. Also, we concentrate on the 
fixed points $\mathcal{C}$ and $\mathcal{D}$ in addition to the Gaussian fixed 
point $\mathcal{A}$; in fact, this turns out to not be a limitation, since 
fixed point $\mathcal{B}$ undergoes a fixed-point collision slightly above 
$\Nf=1$ and hence disappears from the phase diagram anyway. For a more 
transparent analysis, it is useful to introduce the quantity
\begin{equation}
 \Ni = \frac{1}{2^{\lfloor d/2 \rfloor}} \Nf d_\gamma, \label{eq:Nirred}
\end{equation}
which counts the number of irreducible flavor degrees of freedom. Since the 
non-Gaussian fixed points $\mathcal{C}$ and $\mathcal{D}$ lie on their
corresponding coupling axes, the $\beta$ functions reduce to a rather compact 
form,
\begin{widetext}
 \begin{eqnarray}
  \beta_e(e)|_{\kappa=m=0}&=& \frac{\left(-4 e^7-135 \pi ^2 e^5+2025 \pi ^4 e^3\right) N_i-50 \left(5 \pi ^2 e^5-27 \pi ^4 e^3+81 \pi ^6 e\right)}{900 \pi ^4 \left(9 \pi ^2-e^2\right)}, \label{eq:FPD}\\
  \beta_\kappa(\kappa)|_{e=m=0}&=& \frac{8 \left(-184 \kappa ^7+15390 \pi ^2 \kappa ^5+14175 \pi ^4 \kappa ^3\right) N_i+189 \left(176 \pi ^2 \kappa ^5-6210 \pi ^4 \kappa ^3+675 \pi ^6 \kappa \right)}{630 \pi ^2 \left(20 \kappa ^4 N_i+81 \left(5 \pi ^4-2 \pi ^2 \kappa ^2\right)\right)}. \label{eq:FPC}
\end{eqnarray}
\end{widetext}
Solving Eqs.~\eqref{eq:FPD} and \eqref{eq:FPC} for the 
fixed-point condition in the regime where our 
consistency criteria are satisfied yields the fixed point values $e^\ast$ at 
$\mathcal{D}$ and $\kappa^\ast$ at $\mathcal{C}$, respectively, as a 
function of $\Ni$. 
Treating $\Ni$ as a continuous variable for the purpose of illustration, the 
results are shown in Fig.~\ref{fig:EAtDandKappaAtCd3}. In agreement with the 
literature \cite{Pisarski:1984dj,Kubota:2001kk,Kaveh:2004qa,Braun:2014wja}, we 
observe that fixed point $\mathcal{D}$ 
becomes more weakly coupled towards larger flavor numbers. This indicates that 
chiral symmetry is not spontaneously broken, and QED with a large number of 
flavors features a massless conformal long-range phase. The latter is 
quantitatively accessible by means of large-$\Nf$ expansions.

By contrast, the fixed-point value $\kappa^\ast$ at fixed point $\mathcal{C}$ 
increases with flavor number. Moreover, the fixed point disappears in a 
fixed-point collision at $N_{\text{i,cr}}^{\kappa}=5.195$. We conclude that the 
universality class defined by fixed point $\mathcal{C}$ and UV-complete
trajectories emanating from it exist only at small values of $\Ni$. This suggests that 
the universality class, its property of asymptotic safety, as well as possible quantum phase 
transitions induced in the long-range properties are not visible in a 
large-$\Nf$ expansion. We emphasize that the interesting case of $\Nf=2$ and 
$d_\gamma=4$, i.e., $\Ni=4$--discussed in the context of high-$T_\text{c}$ 
cuprate superconductors or graphene--is below the critical flavor number  
$N_{\text{i,cr}}^{\kappa}$. We list our quantitative results for the 
fixed-point properties for this important case in 
Tab.~\ref{tbl:fullFixedPointsd3Nf2dg4}.

\begin{table}
	\centering
\begin{tabular}{ccccccccc}
	& $e$ & $\kappa$ & $m$ & multiplicity & $n_{\mathrm{phys}}$ & 
$\theta_{\mathrm{max}}$ & $\eta_\psi$ & $\eta_\mathrm{A}$ \\
	\noalign{\smallskip} \hline \noalign{\smallskip}
	$\mathcal{A:}$ & $0$ & $0$ & $0$ & $-$ & $2$ & $1.00$ & $0.00$ & $0.00$ \\
	$\mathcal{C:}$ & $0$ & $1.44$ & $0$  & $\mathbb{Z}_2$ & $3$ & $1.05$ & $-0.716$ 
& $0.876$ \\
	$\mathcal{D:}$ & $2.10$ & $0$ & $0$ & $\mathbb{Z}_2$ & $1$ & $1.64$ & $-0.156$ 
& $0.946$ \\
\end{tabular}
\caption{Fixed points of $d=3$ dimensional QED for $\Nf=2$ flavors of reducible 
fermions $d_\gamma=4$, i.e., $\Ni=4$.}
\label{tbl:fullFixedPointsd3Nf2dg4}
\end{table}

The finite range of flavor numbers for which $\mathcal{C}$ exists can accommodate
several scenarios depending on the value of the critical flavor number 
$N_{\text{i,cr}}^\chi$ below which chiral symmetry breaking occurs as 
potentially triggered by a large gauge coupling near fixed point $\mathcal{D}$. 
If  $N_{\text{i,cr}}^\chi<N_{\text{i,cr}}^{\kappa}$, then there is a finite 
window where the system can flow from $\mathcal{C}$ in the UV to $\mathcal{D}$ 
in the IR along a separatrix. Let us for the moment assume that the case 
$\Ni=4$, i.e., $\Nf=2$ for reducible fermions $d_\gamma=4$, is inside this 
conformal window. Then there must be at least one trajectory that connects the 
two fixed points. If the system evolves along this trajectory from the UV to 
the IR, it would constitute an example of emerging chiral symmetry in 
the long-range properties of the theory. The reason is that the UV regime 
characterized by fixed point $\mathcal{C}$ corresponds to a universality class 
without chiral symmetry: the Pauli coupling violates chiral symmetry
explicitly. By contrast, fixed point $\mathcal{D}$ is characterized by 
$m^\ast=0=\kappa^\ast=0$ and thus represents a chirally symmetric universality 
class. 

\begin{figure}[t]
\includegraphics[width=0.45\textwidth]{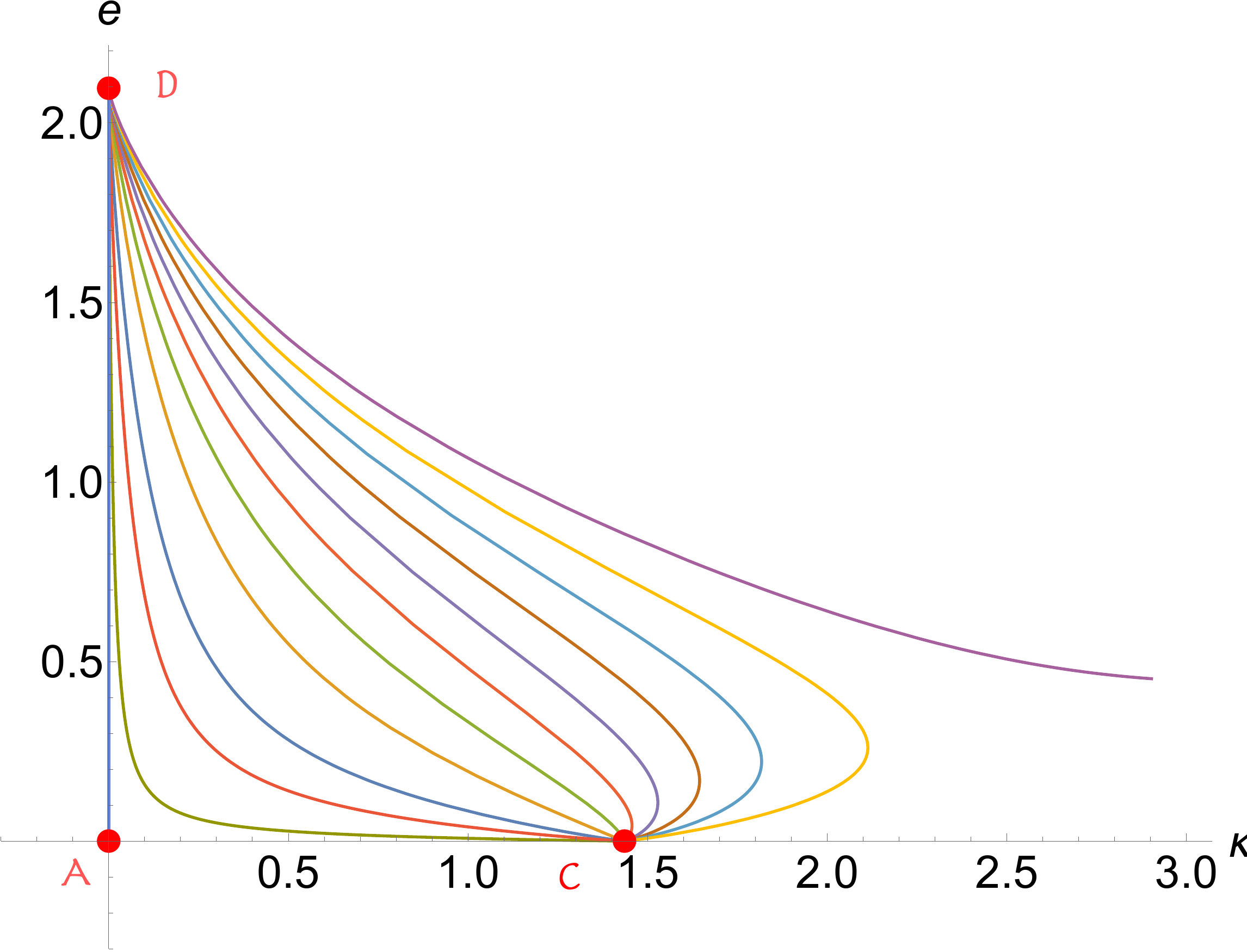}
	\caption{One-parameter family of RG trajectories projected onto the $(\kappa, 
e)$ plane for $N_i = 4$ irreducible flavor degrees of freedom
in $d=3$ spacetime dimensions. These lead from the fully repulsive UV fixed point $\mathcal{C}$ to 
the IR fixed point $\mathcal{D}$ at $\kappa^* = m^* = 0$, thus exhibiting 
emergent chiral symmetry.}
	\label{fig:FamilyOfTrajectoriesCtoD}
\end{figure}

\begin{figure}[t]
\includegraphics[width=0.45\textwidth]{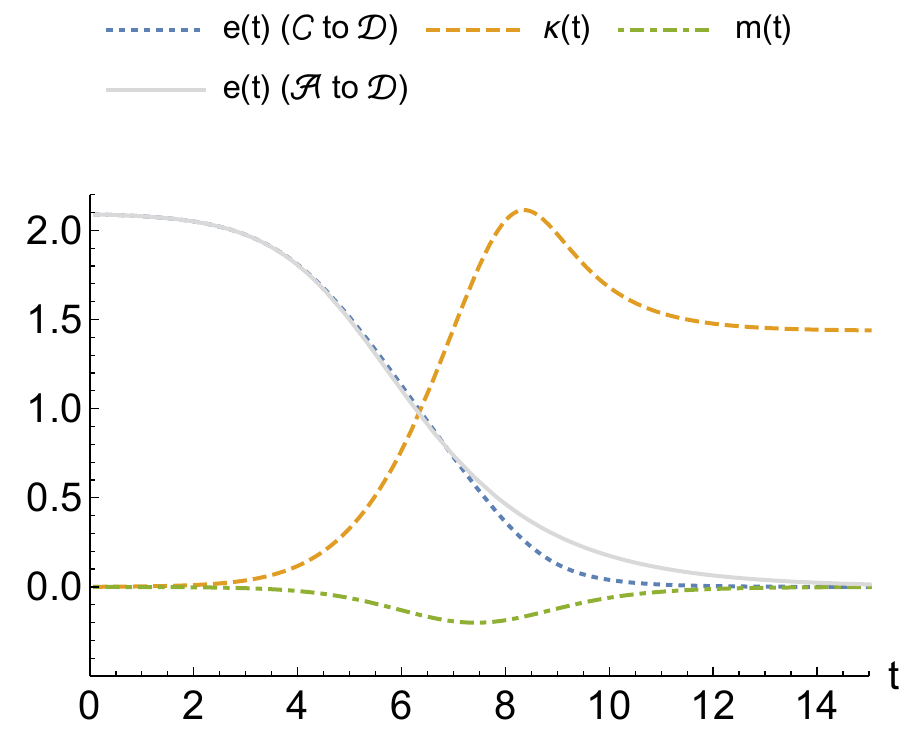}
\caption{RG flow from the fully repulsive UV fixed point $\mathcal{C}$ to the 
IR fixed point $\mathcal{D}$ for $N_i = 4$ irreducible flavor degrees of freedom
in $d=3$ spacetime dimensions; initial conditions are chosen such that the mass 
parameter is zero in the UV and IR. For comparison, the trajectory from the 
Gaussian fixed point $\mathcal{A}$ to $\mathcal{D}$ is also shown (gray).}
	\label{fig:TrajectoryCToD}
\end{figure}

Of course, the mass direction is still a relevant perturbation at fixed 
point $\mathcal{D}$. Since fixed point $\mathcal{D}$ has two irrelevant 
directions in our approximation, a two-dimensional plane of trajectories must 
exist that end exactly in $\mathcal{D}$ and thus in a state of exact chiral 
symmetry. From the UV perspective, fixed point $\mathcal{C}$ has three relevant 
directions; hence, one of them needs to be fine-tuned in order for a trajectory
to end in the chiral plane at fixed point $\mathcal{D}$. A projection of 
this one-parameter family onto the $(\kappa, e)$ plane is depicted in 
Fig.~\ref{fig:FamilyOfTrajectoriesCtoD}. Figure \ref{fig:TrajectoryCToD} further
focuses on the running couplings and mass for a typical trajectory. We observe 
that the transition from the strong-Pauli-coupling regime to the long-range 
value of the gauge coupling coincides with the generation of a finite explicit
mass which nonetheless subsides in the IR. For comparison, we also show an analogous trajectory from the Gaussian fixed point $\mathcal{A}$ to $\mathcal{D}$. 

\section{Pauli-term fixed points in higher dimensions}
\label{sec:d5}

Whereas the search for fixed points in higher dimensions may not offer an 
immediate physical implication, their study illustrates a mechanism underlying 
the existence of the fixed points investigated so far: namely, the competition 
between canonical scaling and quantum fluctuations. Towards higher dimensions, 
the gauge coupling which is marginal in $d=4$ becomes 
power-counting irrelevant, while the power-counting irrelevance of the Pauli 
coupling is further enhanced. The existence of non-Gaussian fixed-points thus 
requires similarly enhanced contributions from quantum fluctuations. If the 
latter contributions are bounded for some reason, we expect the non-Gaussian 
fixed points to disappear towards higher dimensions.

In the present case, we observe that the power-counting irrelevance of the 
gauge coupling $e$ is not counter-balanced by the fluctuation terms which 
further contribute to RG irrelevance. Hence, no fixed point is 
found on the $e$ axis apart from the Gaussian fixed point. In fact, 
slightly above $d=4$ dimensions, the fixed-point structure of $d=4$ as shown 
in Fig.~\ref{fig:StreamPlotAtEEqualTo0d4} persists, but fixed points 
$\mathcal{B}$ and $\mathcal{C}$ approach each other. At a critical dimension 
$d_{\text{cr}}\simeq 4.27$ for $\Nf=1$, the fixed point $\mathcal{C}$ collides 
with fixed point $\mathcal{B}$ (and its $\mathbb{Z}_2$ reflection), such that 
only one fixed point which we call $\mathcal{C}'$ remains on the $\kappa$ 
axis. As a consequence of the fixed-point collision, the new fixed point 
$\mathcal{C}'$ has one relevant direction (and thus one physical parameter) 
fewer than $\mathcal{C}$. Towards higher dimensions, $\mathcal{C}'$ moves 
towards larger values of $\kappa$ and the anomalous dimensions grow beyond 
$\mathcal{O}(1)$.  

Towards larger values of $\Nf$, we observe only quantitative changes, but the 
picture remains qualitatively the same. For instance, for $\Nf=10$, the 
collision of $\mathcal{B}$ and $\mathcal{C}$ occurs at $d_{\text{cr}}\simeq 
4.69$. This number increases beyond $d=5$ for even larger $\Nf$.
The fixed-point value of the Pauli coupling at $\mathcal{C}'$ slightly 
decreases for larger $\Nf$, but the anomalous dimensions remain rather large, 
cf. Tab.\ref{tbl:FixedPointsd5Nfdg4}.

\begin{table}
	\centering
\begin{tabular}{cccccccccc}
	& $ \Nf$ & $e$ & $\kappa$ & $m$ & multiplicity & $n_{\mathrm{phys}}$ & 
$\theta_{\mathrm{max}}$ & $\eta_\psi$ & $\eta_\mathrm{A}$ \\
	\noalign{\smallskip} \hline \noalign{\smallskip}
	$\mathcal{A:}$ & $\forall$  & $0$ & $0$ & $0$ & $-$ & $1$ & $1.00$ & $0.00$ & 
$0.00$ \\
	\textcolor{gray}{$\mathcal{C}'$:} & \textcolor{gray}{1} & 
\textcolor{gray}{$0$} & \textcolor{gray}{$13.9$} & \textcolor{gray}{$0$}  & 
\textcolor{gray}{$\mathbb{Z}_2$} & \textcolor{gray}{$2$} & 
\textcolor{gray}{$5.49$} & \textcolor{gray}{$-2.17$} 
& \textcolor{gray}{$-0.903$}\\
\textcolor{gray}{$\mathcal{C}'$:} & \textcolor{gray}{5} & 
\textcolor{gray}{$0$} & \textcolor{gray}{$10.3$} & \textcolor{gray}{$0$}  & 
\textcolor{gray}{$\mathbb{Z}_2$} & \textcolor{gray}{$2$} & 
\textcolor{gray}{$3.46$} & \textcolor{gray}{$-1.05$} 
& \textcolor{gray}{$-2.14$}\\
	\textcolor{gray}{$\mathcal{C}':$} & \textcolor{gray}{10} & 
\textcolor{gray}{$0$} & \textcolor{gray}{$8.15$} & \textcolor{gray}{$0$}  & 
\textcolor{gray}{$\mathbb{Z}_2$} & \textcolor{gray}{$2$} & 
\textcolor{gray}{$3.24$} & 
\textcolor{gray}{$-0.623$} 
& \textcolor{gray}{$-2.53$} \\
\end{tabular}
\caption{Fixed points of $d=5$ dimensional QED for various flavor numbers of 
irreducible 
fermions $d_\gamma=4$. Fixed point $\mathcal{C}'$ is a remnant of a 
fixed-point collision. In view of the large anomalous dimensions, we interpret 
fixed point $\mathcal{C}'$ as an artifact of the approximation (indicated 
by the gray font).}
\label{tbl:FixedPointsd5Nfdg4}
\end{table}

Since fixed point $\mathcal{C}'$ with its large anomalous dimensions does not 
fully meet our consistency criteria, we consider $\mathcal{C}'$ in 
$d=5$ as an artifact of our approximation. We interpret these findings as 
indicating that an asymptotic-safety scenario for QED induced by the Pauli term may not exist 
in higher dimensions $d=5,6,\dots$. 

\section{Conclusions}
\label{sec:conc}

We have studied the renormalization flow of QED upon the inclusion of a 
Pauli spin-field coupling in general dimensions and flavor numbers. Using the 
functional RG for a nonperturbative estimate of the $\beta$ functions of the 
investigated couplings, we explore the fixed-point structure of the theory 
within a derivative expansion of the effective action. 
We specifically investigate the fate of UV-stable fixed points recently 
discovered in $d=4$ spacetime dimensions for $\Nf=1$ and follow their evolution 
in theory space as a function of the number of dimensions and fermion flavors. 
Such fixed points serve to construct a UV-complete version of QED within an 
asymptotic-safety scenario. They define universality classes that govern the 
physical properties of the theory in the long-range limit. 

Going away from $d=4$ dimensions, we observe the general trend that 
increasing the flavor number tends to destabilize the non-Gaussian fixed points 
discovered in four spacetime dimensions. The most promising candidate for a 
physically relevant universality class is the non-Gaussian 
fixed point at finite Pauli spin-field coupling but vanishing gauge coupling, 
termed fixed point $\mathcal{C}$, which also exists in $d=3$ 
dimensions for sufficiently small flavor 
numbers while satisfying the self-consistency criteria of our approximation. 
In particular, we observe this fixed-point for flavor numbers which are of 
relevance for effective theories of layered condensed-matter systems. This 
universality class may be of interest as it serves as an example where the 
microscopic theory exhibits explicit chiral symmetry breaking, but the 
long-range effective theory may still show a gapless phase protected by an 
emergent chiral symmetry. We explicitly construct RG 
trajectories that emanate from the non-Gaussian 
fixed point $\mathcal{C}$ and approach a long-range regime that is governed by 
the IR-attractive interacting fixed point in the gauge coupling known in the 
QED$_3$ literature. Depending on the flavor number, the latter may correspond 
to a strongly coupled IR phase characterized by spontaneous (or 
dynamical) chiral symmetry breaking, as is subject to an ongoing debate in the 
literature.

It is also interesting to observe that the fixed-point scenario found in $d=4$ 
does not analogously persist above four dimensions. A 
fixed-point collision modifies the phase structure at a critical flavor 
dimension which is in between $d=4$ and $d=5$ for small to moderate flavor 
numbers, but beyond $d=5$ for large flavor numbers. In either case, the fixed 
point observed in $d=5$ no longer meets the quality criteria of our 
approximation, implying that we do not find reliable evidence for a UV 
completion of QED within an asymptotic safety scenario in higher dimensions.   

These findings represent an instructive example for the fact that 
UV completion through asymptotic safety requires a delicate balance of 
dimensional (canonical) scaling and quantum contributions to scaling. Simply 
adding higher-order operators to the truncated effective action is not 
sufficient to induce non-Gaussian fixed points--at least not in the 
computationally controllable part of theory space. This makes the evidence for 
the UV completion of QED exploiting the Pauli coupling as found in 
\cite{Gies:2020xuh} even more remarkable as it is somewhat special to $d=4$. 
It does not analogously exist in higher dimensions, and extends to $d=3$ only 
for small flavor numbers. 

Finally, the important question persists as to whether the strong Pauli 
coupling at the fixed point exerts a relevant influence on higher-dimensional 
operators such as four-fermion interactions. The latter are known to be crucial 
for the status of chiral symmetry in strongly interacting QED both in $d=3$ as 
well as $d=4$ dimensions. This is left for future work. 

\acknowledgments

This work has been funded by the Deutsche Forschungsgemeinschaft (DFG) under 
Grant Nos. 398579334 (Gi328/9-1) and 406116891 within the Research Training 
Group RTG 2522/1.

\appendix
\section{$\beta$ functions}
\label{app:loops}
The quantum contributions to the beta functions along with the anomalous dimensions of the fields were computed in \cite{Gies:2020xuh} and are summarized as follows.
\begin{widetext}
\begin{align}
\label{eq:DeltaBetaE}
\Delta \beta_e
&=
- 4 v_d \frac{\left(d-4\right)\left(d-1\right)}{d} e^3 \; l_d^{(1,\mathrm{B},\tilde{\mathrm{F}}^2)}(0,m^2)
- 16 v_d \frac{\left(d-2\right)\left(d-1\right)}{d} e \kappa^2 \; l_d^{(2,\mathrm{B},\tilde{\mathrm{F}}^2)}(0,m^2) \nonumber\\
&\phantom{=}
- 32 v_d \frac{d-1}{d} e^2 \kappa m \; l_d^{(1,\mathrm{B},\mathrm{F},\tilde{\mathrm{F}})}(0,m^2,m^2) 
- 4 v_d \frac{\left(d-2\right)\left(d-1\right)}{d} e^3 m^2 \; l_d^{(\mathrm{B},\mathrm{F}^2)}(0,m^2) \nonumber\\
&\phantom{=}
- 16 v_d \frac{\left(d-4\right)\left(d-1\right)}{d} e \kappa^2 m^2 \; l_d^{(2,\mathrm{B},\mathrm{F}^2)}(0,m^2) \\
\label{eq:DeltaBetaKappa}
\Delta \beta_\kappa
&=
16 v_d \frac{\left(d-4\right)\left(d-1\right)}{d} \kappa^3 \; l_d^{(2,\mathrm{B},\tilde{\mathrm{F}}^2)}(0,m^2)
- 4 v_d \left( 3 \frac{\left(d-6\right)\left(d-2\right)}{d} + 1 \right) e^2 \kappa \; l_d^{(1,\mathrm{B},\tilde{\mathrm{F}}^2)}(0,m^2) \nonumber\\
&\phantom{=}
+ 4 v_d\,  e^3 m \;
\left[\frac{d-3}{d} \left( l_d^{(1,\mathrm{B},\tilde{\mathrm{F}}_1,\mathrm{F})}(0,m^2,m^2)
-l_d^{(1,\mathrm{B},\mathrm{F}_1,\tilde{\mathrm{F}})}(0,m^2,m^2) \right)
-  \frac{\left(d-4\right)\left(d-1\right)}{2 d} l_d^{(\mathrm{B},\mathrm{F},\tilde{\mathrm{F}})}(0,m^2,m^2) 
\right] \nonumber\\
&\phantom{=}+ 16 v_d\; e \kappa^2 m\; \left[
\frac{5\left(d-4\right)\left(d-3\right)}{2d}  \;l_d^{(1,\mathrm{B},\mathrm{F},\tilde{\mathrm{F}})}(0,m^2,m^2) 
+ \frac{d-3}{d} \;  l_d^{(2,\mathrm{B},\mathrm{F},\tilde{\mathrm{F}}_1)}(0,m^2,m^2)
\right] \nonumber\\
&\phantom{=}
+ 16 v_d\; e \kappa^2 m\; \left[
- \frac{d-3}{d} \;  l_d^{(2,\mathrm{B},\mathrm{F}_1,\tilde{\mathrm{F}})}(0,m^2,m^2)
-  \frac{d+2}{d} \; l_d^{(1,\mathrm{B},\mathrm{F},\tilde{\mathrm{F}})}(0,m^2,m^2)
\right] \nonumber\\
&\phantom{=}
+ 16 v_d \left( 1 - \frac{\left(d-4\right)^2}{d} \right) \kappa^3 m^2 \; l_d^{(1,\mathrm{B},\mathrm{F}^2)}(0,m^2)
+ 4 v_d \frac{\left(d-4\right)\left(d-1\right)}{d} e^2 \kappa m^2 \; l_d^{(\mathrm{B},\mathrm{F}^2)}(0,m^2) \\
\label{eq:DeltaBetaM}
\Delta \beta_m
&=
- 16 v_d \left( d - 1 \right) e \kappa \; l_d^{(1,\mathrm{B},\tilde{\mathrm{F}})}(0,m^2)
+ 16 v_d \left( d - 1 \right) m \kappa^2 \; l_d^{(1,\mathrm{B},\mathrm{F})}(0,m^2)
- 4 v_d \left( d - 1 \right) e^2 m \; l_d^{(\mathrm{B},\mathrm{F})}(0,m^2)
\end{align}
\begin{align}
\eta_\psi &= 
4 v_d \frac{ \left( d - 2 \right) \left( d - 1 \right)}{d}  e^2 \; l_d^{(\mathrm{B},\tilde{\mathrm{F}})}(0,m^2) 
- 8 v_d \frac{d - 1}{d} e^2 \; l_d^{(1,\mathrm{B},\tilde{\mathrm{F}}_1)}(0,m^2) 
+ 16 v_d \frac{\left( d-4 \right) \left( d - 1 \right)}{d} \kappa^2 \; l_d^{(1,\mathrm{B},\tilde{\mathrm{F}})}(0,m^2) \nonumber \\
&\phantom{=} - 32 v_d \frac{d - 1}{d} \kappa^2 \; l_d^{(2,\mathrm{B},\tilde{\mathrm{F}}_1)}(0,m^2) 
+ 32 v_d \frac{d-1}{d} e \kappa m \; l_d^{(1,\mathrm{B},\mathrm{F}_1)}(0,m^2)
\label{eq:etapsi} \\
\eta_A &=
8 v_d \frac{d_\gamma \Nf}{d + 2} e^2 \; l_d^{(2,\tilde{\mathrm{F}}_1^2)}(m^2)
+ 16 v_d d_\gamma \Nf \kappa^2 m^2 l_d^{(\mathrm{F}^2)}(m^2) 
- 16 v_d \frac{d-4}{d} d_\gamma \Nf \kappa^2 \;l_d^{(1,\tilde{\mathrm{F}}^2)}(m^2) \nonumber \\
&\phantom{=} - 64 v_d \frac{d_\gamma \Nf}{d} e \kappa m \; l_d^{(1,\tilde{\mathrm{F}}, \mathrm{F}_1)}(m^2,m^2) 
+ 8 v_d \frac{d_\gamma \Nf}{d} e^2 m^2 \; l_d^{(1,\mathrm{F}_1^2)}(m^2)
\label{eq:etaA}
\end{align}
\end{widetext}
The threshold functions $l_{\dots}^{\dots}$ are defined according to the convention introduced in \cite{Gies:2020xuh}:
\begin{widetext}
\begin{equation}
\label{eq:thresholdNotationDef}
\begin{aligned}
l_d^{([n], X_{[x_d]}^{[x_p]}, Y_{[y_d]}^{[y_p]}, ...)}(\omega_X, \omega_Y, ... ; \eta_X, \eta_Y, ...)
&=
\left(-1\right)^{1 + x_d \, x_p + y_d \, y_p + ...}
\frac{k^{- 2n - d + 2 x_p \left( 1 + x_d \right) + 2 y_p \left( 1 + y_d \right) + ...}}{4 v_d} \\
&\phantom{=}
\times
\int \frac{\mathrm{d}^d p}{\left(2 \pi \right)^d}
\left( p^2 \right)^n \tilde{\partial}_t
\left[ \left( \frac{\partial}{\partial p^2} \right)^{x_d} \Gprop_X(\omega_X) \right]^{x_p}
\left[ \left( \frac{\partial}{\partial p^2} \right)^{y_d} \Gprop_Y(\omega_Y) \right]^{y_p} \dots
\end{aligned}
\end{equation}
\end{widetext}
Parameters in brackets are optional and are understood to have defaults: $n = 0, 
x_d = 0, y_d=0, \dots, x_p = 1,y_p=1, \dots$. 
The sign conventions are such that all threshold functions are positive for 
finite mass parameters $\omega_{X,Y,\dots}$ and vanishing anomalous dimensions 
$\eta_{X,Y,\dots}$.
As is conventional in the literature, the modified scale derivative is 
understood to act only on the regulator terms.
The quantity $\Gprop_X(\omega)$ denotes the inverse regularized propagator of 
type $X$, i.e.,
%
\begin{equation}
\Gprop_\mathrm{B}(\omega) = \frac{1}{P_\mathrm{B} + \omega k^2}, \quad P_{\mathrm{B}} = p^2 \left[ 1 + r_{\mathrm{B}}\left( \frac{p^2}{k^2} \right) \right], 
\end{equation}
\begin{eqnarray}
\Gprop_\mathrm{F}(\omega) &=& \frac{1}{P_\mathrm{F} + \omega k^2}, \quad P_{\mathrm{F}} = p^2 \left[ 1 + r_{\mathrm{F}}\left( \frac{p^2}{k^2} \right) \right]^2\!\!,\\
\Gprop_{\tilde{\mathrm{F}}}(\omega) &=& \frac{1 + r_\mathrm{F}}{P_\mathrm{F} + \omega k^2},
\end{eqnarray}
where $r_{\mathrm{B}}$ and $r_{\mathrm{F}}$ are the boson and fermion regulator shape functions respectively.
More details along with some explicitly computed threshold functions can be 
found in \cite{Gies:2020xuh}.
\bibliography{bibliography}
  
\end{document}